\documentclass[paper,notoc]{JHEP3}

\usepackage{epsfig}
\usepackage{graphicx}
\usepackage{feynmp}

\usepackage{amsmath}
\usepackage{amsfonts}
\usepackage{amssymb}
\usepackage{graphicx}

\def\beq{\begin{equation}}
\def\eeq{\end{equation}}
\newcommand{\bea}{\begin{eqnarray}}
\newcommand{\eea}{\end{eqnarray}}
\newcommand{\nn}{\nonumber}

\def\Eqn#1{Eq.~(\ref{#1})}

\def\bsp#1\esp{\begin{split}#1\end{split}}

\newcommand{\eps}{\epsilon}

\newcommand{\ord}{\begin{cal}O\end{cal}}
\newcommand{\cI}{\begin{cal}I\end{cal}}
\newcommand{\cS}{{\cal S}}
\newcommand{\cC}{{\cal C}}

\newcommand{\rd}{\mathrm{d}}

\def\bit#1\eit{\begin{itemize}#1\end{itemize}}
\def\ben#1\een{\begin{enumerate}#1\end{enumerate}}

\newenvironment{sloppyequation}[0]{\sloppy\begin{flushleft}\hspace*{0.75cm}\(}{\)\end{flushleft}\fussy}
\newenvironment{sloppytext}[0]{\sloppy\begin{flushleft}}{\end{flushleft}\fussy}

\newcommand{\beqsloppy}{\begin{sloppyequation}}
\newcommand{\eeqsloppy}{\end{sloppyequation}}
\newcommand{\btxtsloppy}{\begin{sloppytext}}
\newcommand{\etxtsloppy}{\end{sloppytext}}

\title{The One-Loop One-Mass Hexagon Integral in $\boldsymbol{D=6}$ Dimensions}

\author{Vittorio Del Duca\\
INFN, Laboratori Nazionali Frascati, 00044 Frascati (Roma), Italy\\
Kavli Institute for Theoretical Physics, Santa Barbara\\
       E-mail: \email{delduca@lnf.infn.it}}

\author{Claude Duhr\\
Institute for Particle Physics Phenomenology,
University of Durham\\ Durham, DH1 3LE, U.K.\\
Kavli Institute for Theoretical Physics, Santa Barbara\\
E-mail: \email{claude.duhr@durham.ac.uk}}

\author{Vladimir A. Smirnov\\
Nuclear Physics Institute of Moscow State University\\
Moscow 119992, Russia\\
E-mail: \email{smirnov@theory.sinp.msu.ru}}

\abstract{
We evaluate analytically the one-loop one-mass hexagon in six dimensions.
The result is given in terms of standard polylogarithms of uniform transcendental weight three.}

\keywords{N=4 SYM}
\preprint{IPPP/11/20, DCPT/11/40\\ NSF-KITP-11-070}

\begin{document}

\section{Introduction}
\label{sec:intro}

One-loop scalar $n$-point integrals in $D > 4$ dimensions are interesting objects. Via dimensional shifting, they occur in the $\ord(\eps)$ part of
the one-loop $n$-point MHV amplitude in $D=4-2\eps$ dimensions~\cite{Bern:1996ja,DelDuca:2009ac,Kniehl:2010aj}, and connect
scalar integrals in $(D+2)$ dimensions to tensor integrals in $D$ dimensions~\cite{Bern:1993kr}. Furthermore, it has been noted that they
can be related to higher-loop integrals~\cite{Drummond:2010cz,Anastasiou:2011zk}. If the number of dimensions matches
the number of points, they feature dual conformal invariance~\cite{Drummond:2006rz}, which strongly constrains their form. 
An example is given by the recently computed one-loop scalar massless hexagon integral in $D=6$ 
dimensions~\cite{DelDuca:2011ne,Dixon:2011ng}, whose structure is strikingly similar to the one of the remainder function of two-loop 
amplitudes and Wilson loops~\cite{Bern:2008ap,Drummond:2008aq,DelDuca:2009au,DelDuca:2010zg,Goncharov:2010jf}.
In this contribution, we evaluate analytically the one-loop one-mass hexagon integral in $D=6$ dimensions. The computation is made possible by use of the symbol map~\cite{Goncharov:2010jf}, a certain tensor calculus that allows us to resolve the functional identities among polylogarithms. Using the algorithm of Ref.~\cite{Gangl:2011}, the symbol of the one-loop one-mass hexagon integral in $D=6$ dimensions is then integrated to obtain the analytic expression for the integral.

\section{The one-loop one-mass hexagon integral}
\label{sec:integr}

Let us consider a scalar one-loop one-mass integral in $D=6$ dimensions,
\beq\label{eq:mass_hexagon}
I_{6,m}^{D=6} = \int{\rd^6k\over i\pi^3}\prod_{i=0}^5{1\over D_i}\,,
\eeq
with 
\beq
D_0 = k^2 {\rm~~and~~} D_i = \left(k+\sum_{j=1}^i p_j\right)^2, {\rm~~for~~} i=1,\ldots,5\,,
\eeq
where we have chosen the first momentum as spacelike. Then the mass shell conditions are $p_1^2=m^2 < 0$, and $p_i^2=0$, with
$i=2,\ldots,6$. The momenta are taken all ingoing, such that momentum conservation reads
\beq
\sum_{i=1}^6 p_i=0\,.
\eeq
We consider the integral in Euclidean kinematics where all Mandelstam invariants are taken to be negative, $(p_1+\ldots+p_j)^2<0$, and the integral is real. The one-mass hexagon integral is finite in $D=6$ dimension, so that no regularization is required.
We introduce dual coordinates~\cite{Drummond:2006rz,Kotikov:1991pm,Alday:2007hr},
\beq
p_i = x_i - x_{i+1}\,,
\eeq
with $x_7=x_1$, due to momentum conservation. 

Since the integration measure in Eq.~\eqref{eq:mass_hexagon} is translation invariant, 
we can define $k=x_0-x_1$ and the integral can be rewritten in terms of dual coordinates,
\beq\label{eq:mass_hexagon_X}
I_{6,m}^{D=6} = \int{\rd^6x_0\over i\pi^3}{1\over x_{01}^2\,x_{02}^2\,x_{03}^2\,x_{04}^2\,x_{05}^2\,x_{06}^2}\,,
\eeq
with $x_{ij}^2 = (x_i-x_j)^2 = (p_i+\ldots+p_{j-1})^2$. The mass shell conditions become $x_{12}^2=m^2$ and $x_{23}^2=x_{34}^2=x_{45}^2=x_{56}^2=x_{61}^2=0$. The integral (\ref{eq:mass_hexagon_X}) is invariant under a 
$\mathbb{Z}_2$ symmetry that maps the dual variables as follows,
\beq\label{eq:z2symm}
x_1 \leftrightarrow x_2\,,\quad x_3 \leftrightarrow x_6\,, \quad x_4 \leftrightarrow x_5\,.
\eeq

In Ref.~\cite{Drummond:2006rz} the notion of dual conformal invariance was introduced, \emph{i.e.}, the action of the conformal group on the dual coordinates $x_i$. The integral~\eqref{eq:mass_hexagon_X} transforms covariantly under dual conformal transformations.
A direct consequence of the dual conformal covariance is that $I_{6,m}^{D=6}$ can only depend on dual conformal cross ratios, up to an overall prefactor which carries the conformal weights. For the one-mass six-point kinematics, there are four independent cross ratios, 
given in terms of dual coordinates by
\beq\label{eq:xratios}
u_1 = {x_{26}^2\,x_{35}^2\over x_{25}^2\,x_{36}^2}\,,\,\,\,
u_2 = {x_{13}^2\,x_{46}^2\over x_{36}^2\,x_{14}^2}\,,\,\,\,
u_3 = {x_{15}^2\,x_{24}^2\over x_{14}^2\,x_{25}^2}\,,\,\,\,
u_4 = {x_{12}^2\,x_{36}^2\over x_{13}^2\,x_{26}^2}\,.
\eeq
Under the $\mathbb{Z}_2$ symmetry (\ref{eq:z2symm}), the cross ratios $u_1$ and $u_2$ are exchanged, while $u_3$ and $u_4$ stay
invariant. In terms of the cross ratios (\ref{eq:xratios}), $I_{6,m}^{D=6}$ can be written as
\beq
I_{6,m}^{D=6} = {1\over x_{14}^2\,x_{25}^2\,x_{36}^2}\,\cI_{6,m}(u_1,u_2,u_3,u_4)\,,
\eeq
where the function $\cI_{6,m}$ is manifestly dual conformal invariant,
\beq\label{eq:int6m}
\cI_{6,m}(u_1,u_2,u_3,u_4) = \frac{1}{\sqrt\Delta}\, \cC(u_1,u_2,u_3,u_4)\,,
\eeq
with
\beq\label{eq:delta12}
\Delta = \left( u_1 + u_2 + u_3 - u_1 u_2 u_4 -1 \right)^2 - 4u_1u_2u_3\left(1-u_4\right)\,. 
\eeq
Note that $u_4$ vanishes in the massless limit $x_{12}^2\to 0$, and $\cI_{6,m}$ is reduced to the massless function $\cI_6$
defined in Refs.~\cite{DelDuca:2011ne,Dixon:2011ng}. 

It is easy to derive a Feynman parameter representation for the one-loop one-mass hexagon integral in six dimensions,
\beq
I_{6,m}^{D=6} = \int_0^\infty\left(\prod_{i=1}^6\rd \alpha_i\right)\,\delta\left(1-\sum_{k\in S}\alpha_k\right)\,{2\over \begin{cal}F\end{cal}_{6,m}(\alpha_1,\ldots,\alpha_6)^3}\,,
\eeq
where $\begin{cal}F\end{cal}_{6,m}$ is defined as
\beq
\begin{cal}F\end{cal}_{6,m}(\alpha_1,\ldots,\alpha_6) = \sum_{\stackrel{i,j=1}{i<j}}^6\alpha_i\,\alpha_j\,(-x_{ij}^2)\,,
\eeq
and $S$ denotes a subset of $\{1,\ldots,6\}$. A theorem by Cheng and Wu~\cite{Cheng:1987ga} then guarantees that the Feynman integral is independent of the choice of $S$. In the following we choose $S=\{6\}$, \emph{i.e.}, we freeze the integration variable $\alpha_6$ to $1$. The integrations over $\alpha_4$ and $\alpha_5$ are now trivially performed, leaving us only with a conformally invariant integral to compute,
\beq\bsp
\cI_{6,m}&(u_1,u_2,u_3,u_4) \\
& = \int_0^\infty{\rd \alpha_1\,\rd \alpha_2\,\rd \alpha_3\over (u_2 + \alpha_1 + \alpha_2) (u_4\,\alpha_1 + u_1\, \alpha_3 + \alpha_2) (u_4\,\alpha_1\, \alpha_2 + \alpha_2 + \alpha_1\, \alpha_3 + \alpha_3)}\,.
\esp\eeq
This integral can easily be performed in terms of multiple polylogarithms, leaving us with a rather complicated combination of multiple polylogarithms of weight three. The remarkable simplicity of the massless one-loop hexagon integral in six dimensions~\cite{DelDuca:2011ne,Dixon:2011ng} however suggests that it should be possible to rewrite the result in a much simpler form, a form hidden behind a plethora of complicated functional identities among multiple polylogarithms. These functional identities can be resolved by using the symbol map~\cite{Goncharov:2010jf} which we review in the next section.

\section{The symbol map}
\label{sec:symbol}

The cornerstone of the simplification of the two-loop six-point remainder function~\cite{Bern:2008ap,Drummond:2008aq,DelDuca:2009au,DelDuca:2010zg} is the introduction of the symbol map~\cite{Goncharov:2010jf}, 
a linear map $\begin{cal}S\end{cal}$ that associates a certain tensor to an iterated integral, and thus to a multiple polylogarithm. As an example, the tensor associated to the classical polylogarithm $\textrm{Li}_n(x)$ is,
\beq
\begin{cal}S\end{cal}(\textrm{Li}_n(x)) = -(1-x)\otimes\underbrace{x\otimes\ldots\otimes x}_{(n-1) \textrm{ times}}\,.
\eeq
Furthermore, the tensor maps products that appear inside the tensor product to a sum of tensors,
\beq
\ldots\otimes(x\cdot y)\otimes\ldots = 
\ldots\otimes x\otimes\ldots+
\ldots\otimes y\otimes\ldots\,.
\eeq
It is conjectured that all the functional identities among (multiple) polylogarithms are mapped under the symbol map $\cS$ to algebraic relations among the tensors. Hence, if the symbol map is applied to our expression for $\cI_{6,m}(u_1,u_2,u_3)$, it should capture and resolve all the functional identities among the polylogarithms, and therefore allow us to rewrite the result in a simpler form.

Even though deriving the symbol of the one-loop one-mass hexagon is a rather simple exercise, integrating the symbol back to a function can be much more involved. This can however be achieved by using the algorithm developed in Ref.~\cite{Gangl:2011}, which, after a suitable choice has been made for the functions that should appear in the answer, allows us to reduce the problem of integrating the symbol to a problem of linear algebra.
However, in order to apply this algorithm it is important that all the arguments that enter the tensor be multiplicatively independent. As in our case the arguments of the polylogarithms involve square roots of $\Delta$, this requirement would not be fulfilled. We may remedy this situation
by parametrizing the cross ratios (\ref{eq:xratios}) as
\beq\label{parametrization}
u_1 = {1\over 1-y}\,,\quad u_2 = {v\over v-u}\,,\quad u_3 = {(1-u)(y-x)\over(1-y)(u-v)}\,, \quad u_4 = {v-x\over v}\, ,
\eeq
such that
\beq\label{eq:square12}
\Delta = \frac{(ux-y)^2}{(1-y)^2 (u-v)^2}\, .
\eeq
We note in passing that the Jacobian of the parametrization~\eqref{parametrization} is non zero for generic values of the parameters.

In a nutshell, the algorithm of Ref.~\cite{Gangl:2011} proceeds in two steps:
\begin{enumerate}
\item Given the symbol $\cS(\begin{cal}C\end{cal})$ of the one-loop one-mass hexagon, construct a set of rational functions $\{R_i(u,v,x,y)\}$ such that, \emph{e.g.}, symbols of the form $\cS(\textrm{Li}_n(R_i(u,v,x,y)))$ span the vector space which $\cS(\begin{cal}C\end{cal})$ is an element of.
\item Once a suitable set of rational functions has been obtained, make an ansatz
\beq\bsp
\tilde \cC(u,v,x,y) =&\, \sum_i c_i\,\textrm{Li}_3(R_i(u,v,x,y)) + \sum_{i,j} c_{ij}\,\textrm{Li}_2(R_i(u,v,x,y))\,\ln R_j(u,v,x,y)\\
& + 
\sum_{i,j,k} c_{ijk}\,\ln R_i(u,v,x,y)\,\ln R_j(u,v,x,y)\,\ln R_k(u,v,x,y)\,,
\esp\eeq
where the $c_i$, $c_{ij}$ and $c_{ijk}$ are rational numbers to be determined such that 
\beq
\cS(\tilde \cC) = \cS(\cC)\,. 
\eeq
As the objects appearing in this last equation are tensors (\emph{i.e.}, elements of a vector space), the coefficients $c_i$, $c_{ij}$ and $c_{ijk}$ can equally well be seen as coordinates in a vector space, and the problem of finding the coefficients
 reduces to a problem of linear algebra.
 \end{enumerate}
 We have implemented the algorithm of Ref.~\cite{Gangl:2011} into a {\sc Mathematica} code, which we have applied to the function $\cC(u,v,x,y)$. The result is discussed in the next section.

\section{The one-mass hexagon revealed}
\label{sec:result}

We have found that in the regions where $\Delta$ is negative or where all the $u$'s are smaller than 1,
we can write the function (\ref{eq:int6m}) as
\bea\label{eq:i6m6d}
\lefteqn{ \cI_{6,m}(u_1,u_2,u_3,u_4) } \nn\\
& =  \displaystyle{1\over \sqrt{\Delta}}\,\Bigg[\,-\sum_{i=1}^8\sum_{j=1}^2\left(L_3(x_{i,j}^+,x_{i,j}^-) -{1\over6}\,\bar\ell_1(x^+_{i,j},x^-_{i,j})^3 - {\pi^2\over 6}\,\bar\ell_1(x^+_{i,j},x^-_{i,j})\right)\nn\\
 &\,+{1\over2}\left(\bar\ell_1(x_{2,1}^+,x_{2,1}^-) + \bar\ell_1(x_{2,2}^+,x_{2,2}^-)\right)\left(
 2\bar\ell_1(x_{1,1}^+,x_{1,1}^-)\,\bar\ell_1(x_{1,2}^+,x_{1,2}^-) \right. \nn\\
 &\,\quad+ \bar\ell_1(x_{1,1}^+,x_{1,1}^-)\,\bar\ell_1(x_{3,1}^+,x_{3,1}^-)
 + \bar\ell_1(x_{1,1}^+,x_{1,1}^-)\,\bar\ell_1(x_{3,2}^+,x_{3,2}^-) + \bar\ell_1(x_{1,2}^+,x_{1,2}^-)\,\bar\ell_1(x_{3,1}^+,x_{3,1}^-) \nn\\
&\,\quad\left. + \bar\ell_1(x_{1,2}^+,x_{1,2}^-)\,\bar\ell_1(x_{3,2}^+,x_{3,2}^-) + 2\bar\ell_1(x_{3,1}^+,x_{3,1}^-)\,\bar\ell_1(x_{3,2}^+,x_{3,2}^-)\right)\Bigg]\,,
\eea
where
\bea
L_3(x^+,x^-) &=&\sum_{k=0}^2{(-1)^k\over (2k)!!}\,\ln^k(x^+\,x^-)\,\left(\ell_{3-k}(x^+) - \ell_{3-k}(x^-)\right)\,,\nn\\
\ell_n(x) &=& {1\over 2}\left(\textrm{Li}_n(x) - (-1)^n\textrm{Li}_n(1/x)\right)\,,
\eea
and
\beq
\bar\ell_n(x^+,x^-) = \ell_n(x^+) - \ell_n(x^-)\,.
\eeq
In order to define the arguments of the logarithmic functions, we introduce the variables
\bea
x_{1m}^\pm &=& {u_1+u_2+u_3-u_1u_2u_4 - 1 \pm \sqrt{\Delta}\over2u_1u_2u_3(1-u_4)}\,, \nn\\
\chi^\pm &=&  2u_1u_2u_3(1-u_4)\, x_{1m}^\pm\,.
\eea
%
As functions of the cross ratios, the arguments of the logarithmic functions are,
\bea
x_{1,1}^\pm(u_1,u_2,u_3,u_4) &=& u_3\, x_{1m}^\pm\,, \nn\\
x_{2,1}^\pm(u_1,u_2,u_3,u_4) &=& \frac{(1-u_3)\chi^\pm - 2u_1u_2u_3u_4}{2u_2u_3(1-u_3-u_1u_4)}\,, \nn\\
x_{3,1}^\pm(u_1,u_2,u_3,u_4) &=& \frac{\chi^\pm}{2u_2u_3}\,, \nn\\
x_{4,1}^\pm(u_1,u_2,u_3,u_4) &=& \frac{u_4\left(\chi^\pm(1-u_1u_4)-2u_1u_3(1-u_4)\right)}{2(1-u_4)(1-u_3-u_1u_4)}\,, \nn\\
x_{5,1}^\pm(u_1,u_2,u_3,u_4) &=& \frac{\chi^\pm - 2u_1(1-u_4)}{2u_1u_4(1-u_2)}\,, \nn\\
x_{6,1}^\pm(u_1,u_2,u_3,u_4) &=& \frac{\chi^\pm-2u_1u_2(1-u_4)}{2u_1(1-u_2)(1-u_4)}\,, \nn\\
x_{7,1}^\pm(u_1,u_2,u_3,u_4) &=& \frac{(1-u_1 u_4) \chi ^\pm-2 u_1 u_3 (1-u_4)}{2 u_3 (1-u_1)}\,,\nn\\
x_{8,1}^\pm(u_1,u_2,u_3,u_4) &=& \frac{\chi ^\pm-2 u_1 u_3}{2 u_1 (1-u_3-u_2 u_4)}\,,
\eea
and $x_{i,2}^\pm(u_1,u_2,u_3,u_4)$ are defined from $x_{i,1}^\pm(u_1,u_2,u_3,u_4)$ by exchanging $u_1$ and $u_2$,
\beq
x_{i,2}^\pm(u_1,u_2,u_3,u_4) = x_{i,1}^\pm(u_2,u_1,u_3,u_4)\,, \qquad i=1,\ldots,8\,.
\eeq
Hence, under the $\mathbb{Z}_2$ symmetry, $x_{i,1}^\pm\leftrightarrow x_{i,2}^\pm$, thus making \Eqn{eq:i6m6d}
manifestly symmetric. Furthermore, in the massless limit $u_4\to 0$, we obtain
\bea
&& x_{1,j}^\pm \to x_{1}^\pm\,, \quad x_{2,1}^\pm, x_{3,1}^\pm \to x_2^\pm\,, \quad x_{2,2}^\pm, x_{3,2}^\pm \to x_3^\pm\,, \nn\\
&& x_{4,j}^\pm\to0,\quad x_{5,j}^\pm\to\infty\,, \quad x_{6,1}^\pm \to 1/x_{6,2}^\mp\,, \quad x_{7,j}^\pm\to 1/x_{8,j}^\mp\,,
\eea
with $ j=1,2$, and where the massless hexagon variables are
\beq
x_i^\pm = u_i\,x_{0m}^\pm\,,\quad i=1,2,3\,, \qquad x_{0m}^\pm = {u_1+u_2+u_3-1\pm\sqrt{\Delta_0}\over 2u_1u_2u_3}\,,
\eeq
with $\Delta_0 = (u_1+u_2+u_3-1)^2-4u_1u_2u_3$. Thus
the terms which depend only on $x_{i,j}^\pm$, with $i=1,2,3$, immediately reproduce the massless hexagon. Therefore the contributions from the other variables must vanish. To see that this is the case, we note that 
\bea
L_3(x^+,x^-) &=& - L_3(x^-,x^+) = L_3\left({1\over x^+}, {1\over x^-}\right)\,, \nn\\
 \ell_n(x) &=& (-1)^{n+1}\,\ell_n\left({1\over x}\right)\,.
\eea
Let us take for example the terms in \Eqn{eq:i6m6d} which depend only on $x_{6,j}^\pm$,
\beq\bsp
&L_3(x_{6,1}^+,x_{6,1}^-) -{1\over6}(\ell_1(x^+_{6,1}) - \ell_1(x^-_{6,1}))^3 - {\pi^2\over 6}(\ell_1(x^+_{6,1}) - \ell_1(x^-_{6,1})) \\
+\,&
L_3(x_{6,2}^+,x_{6,2}^-) -{1\over6}(\ell_1(x^+_{6,2}) - \ell_1(x^-_{6,2}))^3 - {\pi^2\over 6}(\ell_1(x^+_{6,2}) - \ell_1(x^-_{6,2}))\,.
\esp\eeq
In the massless limit, this becomes
\beq\bsp
&L_3(x_{6,1}^+,x_{6,1}^-) -{1\over6}(\ell_1(x^+_{6,1}) - \ell_1(x^-_{6,1}))^3 - {\pi^2\over 6}(\ell_1(x^+_{6,1}) - \ell_1(x^-_{6,1})) \\
+\,&
L_3\left({1\over x_{6,1}^-},{1\over x_{6,1}^+}\right) -{1\over6}\left(\ell_1\left({1\over x^-_{6,1}}\right) - \ell_1\left({1\over x^+_{6,1}}\right)\right)^3 - {\pi^2\over 6}\left(\ell_1\left({1\over x^-_{6,1}}\right) - \ell_1\left({1\over x^+_{6,1}}\right)\right)\\
 =\,& 0\,.
\esp\eeq
The same reasoning shows that the terms depending on $x_{7,j}^\pm$ and $x_{8,j}^\pm$ cancel each other. The terms depending on $x_{4,j}^\pm$ and $x_{5,j}^\pm$ are slightly more subtle, because the functions $\ell_n(x)$ are divergent when $x$ approaches either $0$ or $\infty$. Let us concentrate on $x_{4,j}^\pm$. Using the inversion formulae for the polylogarithms,
\beq\bsp
\textrm{Li}_1(x) &\,= \textrm{Li}_1(1/x) +\ln(-x)\,,\\ 
\textrm{Li}_2(x) &\,= -\textrm{Li}_2(1/x) -{1\over2}\ln^2(-x) - {\pi^2\over6}\,,\\ 
\textrm{Li}_3(x) &\,= \textrm{Li}_3(1/x) -{1\over6}\ln^3(-x) - {\pi^2\over6}\ln(-x)\,,
\esp\eeq
we can write the $\ell_n$ functions in the form,
\beq\bsp
\ell_1(x) &\, = \textrm{Li}_1(x)  + {1\over2}\ln(-x)\,,\\
\ell_2(x) &\, = \textrm{Li}_2(x)  + {1\over4}\ln^2(-x) + {\pi^2\over12}\,,\\
\ell_3(x) &\, = \textrm{Li}_3(x)  + {1\over12}\ln^3(-x) + {\pi^2\over12}\ln(-x)\,.
\esp\eeq
In the limit $x\to0$, the $\ell_n$ function splits into two pieces, a polylogarithmic piece that vanishes powerlike and a logarithmically divergent piece. A little algebra then shows that the logarithms conspire in such a way that
\beq
\lim_{u_4\to 0}\left(L_3(x_{4,j}^+,x_{4,j}^-) -{1\over6}(\ell_1(x^+_{4,j}) - \ell_1(x^-_{4,j}))^3 - {\pi^2\over 6}(\ell_1(x^+_{4,j}) - \ell_1(x^-_{4,j})) \right)=0\,.
\eeq
The same reasoning of course applies to $x_{5,j}^\pm$. Thus, in the massless limit $u_4\to 0$, \Eqn{eq:i6m6d} reduces to the massless 
hexagon~\cite{DelDuca:2011ne,Dixon:2011ng}.

We stress that \Eqn{eq:i6m6d} is valid only in the regions where $\Delta < 0$, or where all the $u$'s are smaller than 1.
Outside those regions, the analytic structure of \Eqn{eq:i6m6d} seems to be more complicated. We plan to study that in the near future.

\section{Conclusions}
\label{sec:concl}

In this paper, we have continued the exploration undertaken in Ref.~\cite{DelDuca:2011ne}, and we have computed
analytically the one-loop one-mass hexagon integral in $D=6$ dimensions. Just like for the massless hexagon integral,
the result is given in terms of standard polylogarithms of uniform transcendental weight three, and in the massless limit
it reduces manifestly to the massless hexagon. The similarity in structure between the massless and 
one-mass hexagons, coupled with the similarity between the one-loop massless hexagon in $D=6$ dimensions and the remainder function of
the two-loop hexagon Wilson loops and amplitudes in $D=4$ dimensions, points to a similar simple structure for hexagons with more masses
in $D=6$ dimensions and for Wilson loops and amplitudes with 7 or more points in $D=4$ dimensions. This is left for future work.

\section*{Acknowledgements}
VDD and CD are grateful to the KITP, Santa Barbara, for the hospitality at the later stages of this work. CD is grateful to Herbert Gangl for valuable discussions on the symbol technique. 
This work was partly supported by the Research Executive Agency (REA) of the European Union through the Initial Training Network LHCPhenoNet under contract PITN-GA-2010-264564, by the Russian Foundation for Basic Research through grant 11-02-01196 and by the National Science Foundation under Grant No. NSF PHY05-51164.

\end{document}